\documentclass[aps,prl,reprint,groupedaddress]{revtex4-1}

\usepackage{amsmath}
\usepackage{amssymb}
\usepackage{amsfonts}
\usepackage{amsbsy}
\usepackage{bm}
\usepackage{color}
\usepackage{array}
\usepackage{dcolumn}
\usepackage[colorlinks=true, allcolors=blue]{hyperref}
\usepackage{colortbl}
\usepackage{tensor}
\usepackage{braket}
\usepackage{graphicx}
\usepackage{eufrak}
\usepackage{mathrsfs}
\usepackage{slashed}

\newcommand{\mcL}{\mathcal{L}}

\newcommand{\FuC}[2]{\tensor{\mathcal{F}}{^{{#1} {#2}}}}
\newcommand{\FdC}[2]{\tensor{\mathcal{F}}{_{{#1} {#2}}}}
\newcommand{\FudC}[2]{\tensor{\mathcal{F}}{^{#1}_{#2}}}

\newcommand{\FuCt}[2]{\tensor{\widetilde{\mathcal{F}}}{^{{#1} {#2}}}}

\newcommand{\FudCt}[2]{\tensor{\widetilde{\mathcal{F}}}{^{#1}_{#2}}}

\newcommand{\fu}[2]{\tensor{f}{^{{#1} {#2}}}}

\newcommand{\fd}[2]{\tensor{f}{_{{#1} {#2}}}}

\newcommand{\Fu}[2]{\tensor{F}{^{{#1} {#2}}}}
\newcommand{\Fd}[2]{\tensor{F}{_{{#1} {#2}}}}

\newcommand{\fut}[2]{\tensor{\widetilde{f}}{^{{#1} {#2}}}}

\newcommand{\Fut}[2]{\tensor{\widetilde{F}}{^{{#1} {#2}}}}

\newcommand{\Hu}[2]{\tensor{H}{^{{#1} {#2}}}}

\newcommand{\mcX}{\mathcal{X}}
\newcommand{\mcY}{\mathcal{Y}}

\begin{document}


\title{Cherenkov radiation from the quantum vacuum}


\author{Alexander J. Macleod}
\author{Adam Noble}
\email{adam.noble@strath.ac.uk}

\author{Dino A. Jaroszynski}
\affiliation{SUPA Department of Physics, University of Strathclyde, Glasgow G4 0NG, United Kingdom}


\date{\today}

\begin{abstract}
A charged particle moving through a medium emits Cherenkov radiation when its velocity exceeds the phase velocity of light in that medium. Under the influence of a strong electromagnetic field, quantum fluctuations can become polarized, imbuing the vacuum with an effective anisotropic refractive index and allowing the possibility of Cherenkov radiation from the quantum vacuum. We analyze the properties of this vacuum Cherenkov radiation in strong laser pulses and the magnetic field around a pulsar, finding regimes in which it is the dominant radiation mechanism. This radiation process may be relevant to the excess signals of high energy photons in astrophysical observations.
\end{abstract}


\maketitle

Quantum electrodynamics (QED) is one of the most successful and well tested theories in physics. An early prediction of QED is the presence of virtual particle-antiparticle pairs which fluctuate in and out of existence in the quantum vacuum. It has been known since the seminal work of Euler and Heisenberg \cite{heisenberg1936} (see also \cite{schwinger1951}) that a strong electromagnetic field can polarize these vacuum fluctuations. This, in turn, can mediate an indirect interaction between a probe photon and the strong field such that the photon propagates as it would in a dielectric medium (for extensive reviews see \cite{marklund2006nonlinear,di2012extremely,battesti2012magnetic} and references therein). Euler-Heisenberg theory is the result of integrating out the fermion degrees of freedom in the QED path integral, producing a nonlinear effective theory in which the photon interacts directly with the strong field. Other nonlinear theories of electrodynamics have been proposed, most notably Born-Infeld theory \cite{born1934foundations}, which in its original form was an attempt to resolve the electron self-energy problem before the advent of QED. More recently, it has found a resurgence of interest due to its emergence in the low energy limit of some string theories \cite{fradkin1985,abalos2015}.

It is well known that a charged particle moving through a material medium can emit Cherenkov radiation \cite{cherenkov1934visible,vavilov1934ossible}. The first theoretical work to explain these results was presented by Frank and Tamm \cite{tamm1937coherent} (though earlier work by Heaviside \cite{heaviside1888electro} and Sommerfeld \cite{sommerfeld1904simplified} considered similar effects). This effect occurs because, in a medium with refractive index $n$, the phase velocity of light is reduced, $v_p = c/n$, so a particle traveling through the medium with velocity $V > v_p$ will outrun any electromagnetic waves it emits. This can lead to the emission of radiation due to the build up of wavefronts propagating from the particle, producing the well known ``Cherenkov cone'' of radiation behind the particle.

Since vacuum fluctuations can also reduce the phase velocity of light (see for example  \cite{flood2013}), the same argument implies that high-energy particles traveling through strong electromagnetic fields should emit Cherenkov radiation, in addition to the usual synchrotron radiation caused by acceleration in the field. First steps towards analyzing this effect were taken by Erber \cite{erber1966high}, who used the principles of QED to obtain semi-quantitative predictions for the radiation emitted by an electron in a strong magnetic field. This was followed by Ritus \cite{ritus1985quantum}, who derived the analogous process for an electron in constant crossed fields from the effective photon mass. Subsequently Dremin \cite{dremin2002cherenkov} made more quantitative estimates for the Cherenkov radiation produced by particles crossing a laser pulse, while Marklund {\em et al.} \cite{marklund2005cherenkov} explored the possibility of Cherenkov radiation from a particle in a photon gas.

In this Letter, we provide a unified description of vacuum Cherenkov radiation in nonlinear electrodynamics, applicable to arbitrary field configurations. To illustrate the approach we analyze the effect in the context of both upcoming laser facilities (e.g. the Extreme Light Infrastructure (ELI) \cite{eli}) and astrophysical sources of strong fields. In the latter, we find regimes in which the Cherenkov radiation dominates over other radiation processes, highlighting a new and as yet unexplored mechanism for generating gamma rays, which we suggest should be further investigated in the context of the observed excess signals of astrophysical high energy photons \cite{fermi-LAT2017,abdo2007,hooper2011,gordon2015}.

Lorentz invariance of the vacuum requires nonlinear theories of electrodynamics to be constructed from Lagrangians, $\mcL(X,Y)$, depending only on the two electromagnetic invariants, $X=-\frac{1}{4}\Fu{\mu}{\nu}\Fd{\mu}{\nu}$ and $Y=-\frac{1}{4}\Fut{\mu}{\nu}\Fd{\mu}{\nu}$, where $\Fu{\mu}{\nu}$ and $\Fut{\mu}{\nu}$ are the electromagnetic field and its dual, and repeated indices imply summation. Given the success of Maxwell's theory, we consider only leading order corrections, i.e., Lagrangians of the form
\begin{equation}
\label{Lagrangian}
{\cal L}= X + \lambda_+ X^2+ \lambda_- Y^2,
\end{equation}
where the constants $\lambda_{\pm}$ determine the specific theory. For Euler-Heisenberg, 
\begin{equation}
\label{EH_Lagrangian}
\frac{1}{4}\lambda_+
=
\frac{1}{7}\lambda_-
=
\frac{\alpha}{90\pi}\frac{1}{E^2_S}
,
\end{equation}
where $\alpha\simeq 1/137$ is the fine-structure constant and $E_S=m^2_e/e\simeq1.3\times 10^{18}$ V/m is the Schwinger field~\cite{schwinger1951} (we work throughout in units where $c=\hbar=1$). In Born-Infeld theory the (unknown) constants coincide, $\lambda_+ = \lambda_-$~\cite{born1934foundations}. Although our results are readily extendible to more general Lagrangians, (\ref{Lagrangian},\ref{EH_Lagrangian}) remains a good approximation for field strengths approaching $E_S$, and so is sufficient for our purposes.

The field equations following from (\ref{Lagrangian}) are $\partial_\mu \Fut{\mu}{\nu}=0$ and  $\partial_\mu \Hu{\mu}{\nu}=0 $, where the excitation tensor $\Hu{\mu}{\nu}=(1+2\lambda_+ X)\Fu{\mu}{\nu}+2\lambda_- Y\Fut{\mu}{\nu}$. Taking $\Fu{\mu}{\nu}$ to be the sum of a strong, slowly varying background $\FuC{\mu}{\nu}$ and a weaker radiation field $\fu{\mu}{\nu}$, and linearizing in the latter, yields
\begin{equation}
\label{perteqn}
\partial_\mu \fut{\mu}{\nu}=0, \qquad \partial_\mu \left(\chi^{\mu\nu\alpha\beta}\fd{\alpha}{\beta}\right)=0,
\end{equation}
with the constitutive tensor
\begin{align}
\chi^{\mu\nu\alpha\beta}
=&
(1+2\lambda_+ \mcX)
(
g^{\mu\alpha}g^{\nu\beta}
-
g^{\mu\beta}g^{\nu\alpha}
)
\nonumber\\
&
-
2\lambda_+ \FuC{\mu}{\nu}\FuC{\alpha}{\beta}
-
2\lambda_-\FuCt{\mu}{\nu}\FuCt{\alpha}{\beta}.
\label{constitutive}
\end{align}
$g^{\mu\nu}$ is the metric tensor, and we define $\mcX=-\frac{1}{4}\FuC{\mu}{\nu}\FdC{\mu}{\nu}$, and similarly for $\mcY$.

The system (\ref{perteqn},\ref{constitutive}) has been well studied (e.g. \cite{bialynicka1970nonlinear,obukhov2002fresnel}). Neglecting derivatives of the background~\footnote{This restricts our results to backgrounds that vary slowly on the scale of the radiation. However, since we are primarily interested in radiation of extremely short wavelengths, in practice this includes all backgrounds one might want to consider.}, and defining the phase $\varphi = k_\mu x^\mu$, the radiation field can be expressed as $\fd{\mu}{\nu}=(k_\mu a_\nu-k_\nu a_\mu)e^{i\varphi}$, where the polarization $a_\mu$ and wavevector $k_\mu$ are determined algebraically from
\begin{equation}
\chi{^{\mu\nu\alpha\beta}}
k_\nu
k_\beta
a_\alpha
=0.
\label{algebraic}
\end{equation}
This has solutions $a^\mu_+ = \FudC{\mu}{\nu} k^\nu_+$, $a^\mu_-=\FudCt{\mu}{\nu}k^\nu_-$ \cite{bialynicka1970nonlinear}, with the wavevectors obeying the dispersion relations 
$
k^2_\pm
\simeq
2 \lambda_\pm
\FdC{\lambda}{\mu}
\FudC{\lambda}{\nu}
k^\mu_\pm
k^\nu_\pm
$, where only the leading order behaviour in $\lambda_{\pm}$ has been included. Evidently, for  $\lambda_+ \ne \lambda_-$ we have birefringence. Using $k^2 = (\omega^2 - |{\bf k}|^2) = (v_p^2 - 1)|{\bf k}|^2$, the dispersion relations yield the phase-velocity $v_p$,
\begin{align}
v_{p \pm}^2
\simeq&
1
+
2 \lambda_\pm
\FdC{\lambda}{\mu}
\FudC{\lambda}{\nu}
\hat{k}^\mu_\pm
\hat{k}^\nu_\pm
,
\label{phase}
\end{align}
where $\hat{k}_\pm^\mu = k_\pm^\mu/|{\bf k_\pm}|$ is the direction 4-vector of the radiation. For Euler-Heisenberg, the last term in (\ref{phase}) encodes the effect of the photon mass operator in QED \cite{erber1966high,ritus1985quantum}.

To interpret these solutions as Cherenkov radiation, we must relate them to the source of the radiation---i.e., the charged particle. The analogous problem in a material medium has been well-studied~\cite{tamm1937coherent}, and leads to the well-known expressions for the emission angle $\theta_C$ relative to the particle's velocity, and the power radiated per unit frequency $dP/d\omega$,
\begin{equation}
\cos \theta_C
=
\frac{v_p}{|\bm{\beta}|}
,
\qquad
\frac{d P}{d\omega}
=
\frac{e^2}{4\pi}
\omega
\sin^2 \theta_C
\label{Frank}
.
\end{equation}
These are calculated for Cherenkov radiation in a homogeneous, isotropic medium (ICR). 
Each of the expressions~(\ref{Frank}) are relatively simple in their structure, and it can clearly be seen that the key parameters are the phase velocity $v_p$, and the particle velocity $\bm{\beta}$. The definition of the Cherenkov angle ensures that no Cherenkov radiation is observed for $\beta< v_p$, ($\beta \equiv |\boldsymbol{\beta}|$). 

The generalization of the Cherenkov angle to nonlinear electrodynamics is straightforward: it retains the form given in (\ref{Frank}), but the anisotropy of the background field implies the phase velocity itself depends on the direction of emission, $v_p=v_p(\hat k)$. This can be accounted for using~(\ref{phase}):
\begin{equation}
\cos^2\theta^\pm_C
=
\frac{1}{\beta^2}
\left(
1
+
2\lambda_{\pm}
\FdC{\lambda}{\mu}
\FudC{\lambda}{\nu}
\hat{k}^\mu_\pm
\hat{k}^\nu_\pm
\right)
\label{AngleNLVE}
,
\end{equation}
which is valid in any (slowly varying) background field. Note that, in theories exhibiting birefringence, we have two Cherenkov cones, corresponding to the different phase velocities of the two polarizations.

It is clear that the Cherenkov power formula in (\ref{Frank}) cannot be adopted directly into the nonlinear theories as in general $\theta^\pm_C$ depends on the azimuthal angle, and we must instead determine the differential power emitted per unit frequency {\em per unit azimuthal angle}, $d^2P/d\omega d\phi$. 
We follow the approach taken by Altschul  to describe Cherenkov radiation in Lorentz-violating vacua~\cite{altschul2007cerenkov}. Although the physical basis of such theories is the reverse of nonlinear electrodynamics (where Lorentz invariance is strictly preserved), the linearization treats the background field as an external structure, and (\ref{perteqn}) is formally equivalent to CPT-even Lorentz-violating electrodynamics. The key observation is that Cherenkov modes corresponding to different wavevectors $k^\mu$ propagate independently, and hence behave as waves propagating in an isotropic medium with scalar refractive index $n=1/v_p(\hat{k})$. ICR is linearly polarized in the plane $(\bm{\hat{\beta}},{\bf \hat{k}})$, and orthogonal to $\bf{\hat{k}}$, i.e., in the direction $\bm{\hat{\epsilon}_0}=(\bm{\hat{\beta}}-\cos\theta_C {\bf \hat{k}})/\sin\theta_C$ (throughout, spatial vectors with carets are unit normalized). In the nonlinear case there are two independent polarization modes, $a^\mu_+$ and $a^\mu_-$, the spatial parts of which do not in general coincide with $\bm{\hat{\epsilon}_0}$. As such, only the projection of ICR along these directions will propagate:
\begin{align}
\frac{d^2 P_{\pm}}{d\omega d\phi}
=&
\frac{e^2}{8\pi^2}
|
\bm{\hat{\epsilon}_0}
.
\bm{\hat{\epsilon}_\pm}
|^2
\omega
\sin^2 \theta^{\pm}_C
.
\label{power}
\end{align}
Here, $\bm{\hat{\epsilon}_\pm}$ are the (unit normalized) spatial components of the polarization modes $a^\mu_\pm$ (see Appendix~A for details). 
The derivation of (\ref{power}) treats the particle's orbit as rectilinear. It is therefore valid only for wavelengths that the particle can emit while turning a negligible angle. See Appendix~B for a demonstration that this includes almost all the radiation in the examples below.

As can be seen from (\ref{Frank}) and (\ref{power}), the Cherenkov spectrum has an explicit linear dependence on the frequency $\omega$. This means the spectrum appears to diverge at high frequencies. In a material medium, dispersive effects give $\theta_C$ an $\omega$ dependence, so that at high frequencies Cherenkov radiation is suppressed. In nonlinear electrodynamics, this is not the case, and we must assume a cut-off frequency will arise from physics not captured in $\mcL(X,Y)$ (see \cite{altschul2007vacuum} for an analogous discussion in the context of Lorentz-violating electrodynamics). In the case of QED, for example, the Euler-Heisenberg Lagrangian must be supplemented by higher derivative terms at very high frequencies 
\cite{marklund2006nonlinear}
. We could simply impose a cut-off directly on the frequency $\omega$. However, since frequency is not a Lorentz invariant  this would not be a physically meaningful condition. Instead, we assume~(\ref{power}) is valid for photons with small quantum non-linearity parameter~\cite{ritus1985quantum},
\begin{equation}
\chi_{\gamma}
=
\frac{|e|}{m^3_e}
\sqrt{-\FuC{\mu}{\lambda}
\FudC{\nu}{\lambda}
 k_\mu k_\nu}
\lesssim 1,
\label{cut}
\end{equation}
which can be solved for the maximum frequency, $\omega_\text{max}$. This is not strictly a cut-off, and Cherenkov radiation may still occur for higher frequencies, but our results may not be reliable above $\omega_\text{max}$.

In principle, $\gamma$ should reduce as the particle loses energy to radiation. This could be accounted for by introducing a damping force ${\bf F}_d=-{\cal P}_C\bm{\beta}$, where ${\cal P}_C$ is the integral of (\ref{power}), and solving simultaneously for the motion of the particle and the radiation. In practice, however, this is generally unnecessary, as for $\gamma\gg 1/\sqrt{1-v^2_p}$ we can set $\beta=1$ in (\ref{AngleNLVE},\ref{power}).

With these considerations, we now have all the ingredients necessary to determine the Cherenkov radiation emitted by a particle moving in any given field configuration. To demonstrate more concretely the vacuum Cherenkov effect, we consider two examples of field configurations: a constant crossed field (representing a laser pulse) and a constant magnetic field (representing the field around a pulsar).

Advances in laser technology have begun to provide a platform to study strong field effects experimentally, for example the recent results concerning radiation reaction \cite{cole2018experimentallong,poder2018experimentallong}. Many results pertaining to strong field physics locally approximate the laser beam as a constant crossed field. We will consider the background field $\FuC{\mu}{\nu}$ to represent a constant crossed field (i.e., $\mcX=\mcY=0$) of strength $E$, with Poynting vector in the $\bf {\hat{z}}$-direction. 
We consider the electron to be counter-propagating with respect to the Poynting vector, as in this configuration the energy transfer between background and electron will be greatest, leading to the strongest effect. With the set up described the phase velocity can be determined via (\ref{phase}), which leads to the simple expression for the Cherenkov angles, 
\begin{align}
\cos^2 \theta_C^\pm
=&
\frac{
1
}{
\left[
\beta^2
+
2
\left(
1+\beta
\right)^2
\lambda_{\pm}
E^2
\right]
}
.
\label{crossedangle}
\end{align}
In this case the Cherenkov angles are independent of the azimuthal angle $\phi$, and yield 
the Cherenkov condition 
$
(
\gamma
+
\sqrt{\gamma^2 -1}
)^2
E^2
>
1/2\lambda_{\pm}
$. Cherenkov radiation will occur whenever this condition is satisfied, however to be observable it must be non-negligible in comparison with the emission of synchrotron radiation by electrons oscillating in the field. The synchrotron spectrum is \cite{schwinger1949classical}
\begin{align}
\frac{d P_{\text{Synch}}}{d\omega}
=&
\frac{\sqrt{3}}{\pi}
\frac{e^3 E}{m_e c^3}
\frac{\omega}{\omega_c}
\int_{\omega/\omega_c}^{\infty}
dx
K_{5/3}(x)
,
\label{CrossedSynch}
\end{align}
where $K_\nu(x)$ is the order $\nu$ modified Bessel function of the second kind and $\omega_c=3\frac{e E}{m_e c}\gamma^2$. To compare the two radiation processes, we integrate the Cherenkov spectrum with respect to azimuthal angle, and consider the total power per unit frequency,
\begin{align}
&
\frac{d P_{\text{Cher}}}{d\omega}
=
\int_{0}^{2\pi}
d\phi
\left(
\frac{d^2 P_{+}}{d\omega d\phi}
+
\frac{d^2 P_{-}}{d\omega d\phi}
\right)
,
\label{PowerTotal}
\end{align}
where the contribution from each individual mode is determined by (\ref{power}).

\begin{figure}
\centering
    \includegraphics[width=0.4\textwidth,{trim= 0 0 0.7cm 0},clip]{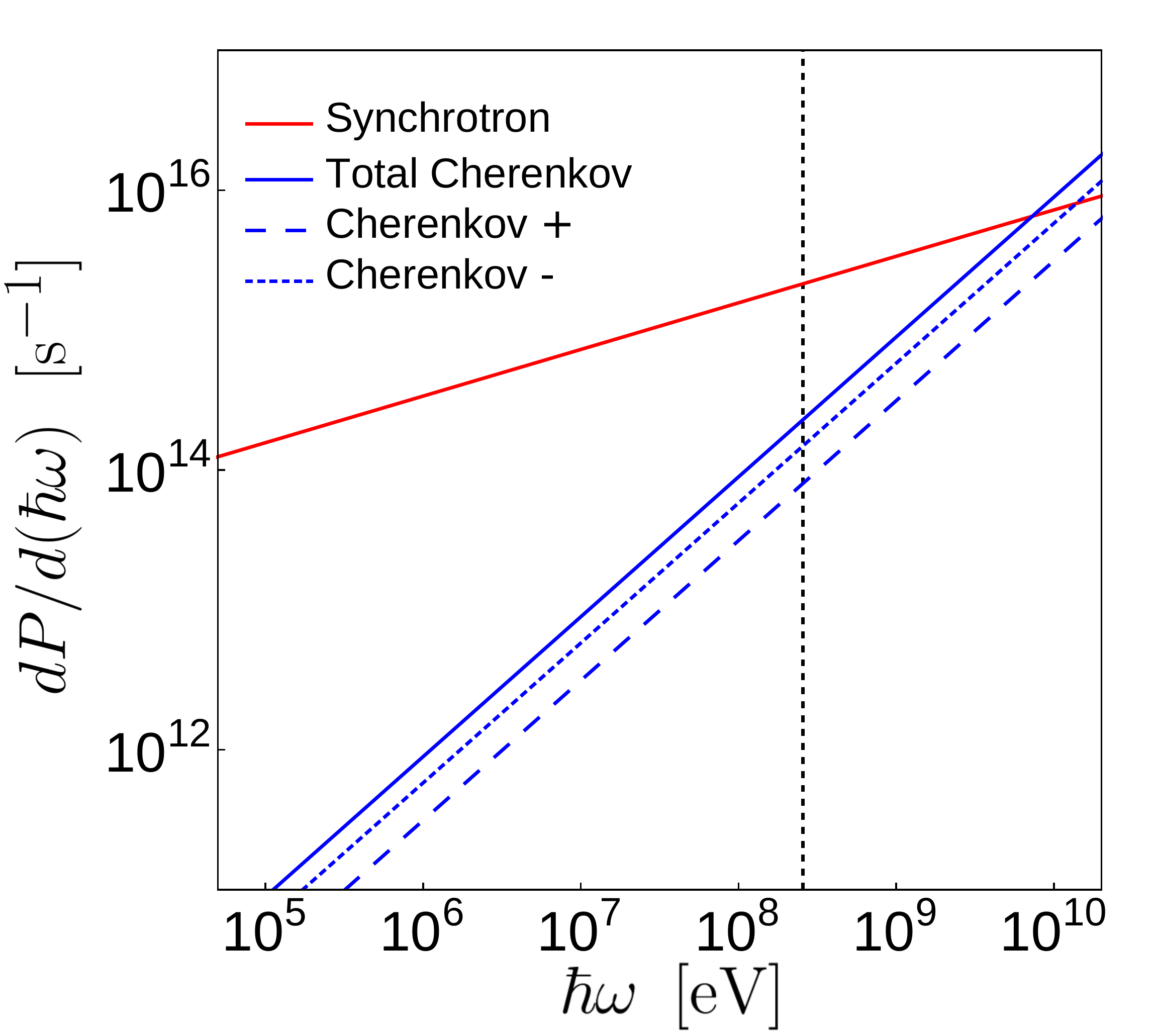}
\caption{ 
Radiated power from the interaction of an electron with $\gamma = 10^5$ and a crossed field with field strength $E = E_S \times 10^{-3}$, due to: Synchrotron radiation (red); total Cherenkov radiation (blue, solid); Cherenkov $+$ mode (blue, dashed); Cherenkov $-$ mode (blue, dot-dashed). The cut-off energy (black, dashed) is $\hbar \omega_{\text{max}} \simeq 0.25$ GeV.}
\label{fig:Crossed}
\end{figure}

Future laser facilities such as ELI \cite{eli} are expected to reach field strengths on the order of $E \sim E_S \times 10^{-3}$, with access to electrons up to $\gamma \sim 10^5$ ($\simeq 50$~GeV). Thus, we consider this parameter regime in comparing the spectra from Cherenkov and synchrotron radiation in a constant crossed field. We also specialize to the Euler-Heisenberg Lagrangian (\ref{EH_Lagrangian}), as this represents arguably the best motivated nonlinear extension to Maxwell electrodynamics. Figure \ref{fig:Crossed} shows the calculated power per unit frequency due to each of the radiation processes as a function of the emitted photon energy $\hbar \omega$. The black dashed line represents the cut-off found from~(\ref{cut}), $\hbar \omega_{\text{max}} \sim (m_e^3 c^5)/(2 e \hbar E) \simeq 0.25$~GeV. Below this limit, synchrotron radiation is always the dominant process. Thus, observing the Cherenkov effect appears unlikely for even future laser facilities. This is primarily due to the limitation on the ability to produce high energy electrons in the lab. For $\gamma \gg 1$, the Cherenkov spectrum becomes proportional to $E^2$, to leading order, and so for a fixed field strength, increasing the energy of the particles has very little effect on the Cherenkov spectrum. Conversely, the synchrotron spectrum becomes increasingly suppressed as $\gamma$ increases for fixed $E$. For the field strength considered here, 
 an electron Lorentz factor $\gamma \sim 2.5 \times 10^6$, corresponding to an energy of $1.3$~TeV, would be required to have the contributions from Cherenkov and synchrotron processes approximately equal at the cut-off
. There is also the concern that to reach these high field strengths in a real experiment, strong focussing techniques are needed to compress the laser pulse, and this brings in a significant range of other effects which would act to drown out the Cherenkov signal, or deplete the electron energy sufficiently that, by the time it reaches the peak intensity of the pulse, its energy has fallen below the Cherenkov threshold \cite{kravets2013}. It might be hoped that protons offer a viable alternative, since their mass greatly suppresses synchrotron radiation. However, the Cherenkov threshold corresponds to a proton energy of $33$ TeV, well beyond what can currently be produced. The possibility of observing Cherenkov radiation in this context therefore seems bleak.

Since the main obstacle to observing Cherenkov radiation is the availability of high energy particles, it is natural to turn our attention to astrophysics, where the only limit on the particle energy is the so-called GZK limit, $\gamma \lesssim 10^{11}$ \cite{greisen1966end,zatsepin1966upper}. Astrophysical objects such as pulsars have also been observed to generate magnetic fields up to and exceeding the Schwinger magnetic field $B_S = E_S/c \simeq 4.4 \times 10^{9}$ T. This makes such a scenario ideal for the study of nonlinear vacuum Cherenkov radiation. As such, we consider a constant magnetic field of strength $B$. We take the particle's velocity perpendicular to the field, since any parallel component of $\bm{\beta}$ can be removed by a Lorentz transformation which does not alter the form of the background field. Taking the $\bf{\hat{z}}$-axis along the particle's velocity, the polar angle of the emitted radiation and the Cherenkov angle coincide, $\theta = \theta_C$. This immediately gives an azimuthal dependence to the Cherenkov angle,
\begin{align}
{\cos^2 \theta_{C}^{\pm}}
=&
\frac{
1
-
2\lambda_{\pm}
B^2
\cos^2 \phi
}{
\beta^2
+
2\lambda_{\pm}
B^2
\sin^2 \phi
}
\label{AngleNLVEMagPerp}
,
\end{align}
which gives the Cherenkov condition,
$\gamma^2B^2>1/2 \lambda_{\pm}$.

We again need to compare the Cherenkov and synchrotron spectra. In the case of the magnetic field we use (\ref{CrossedSynch}), with the substitution $E \rightarrow B c/2$ (the factor 2 arises because the constant crossed field has both magnetic and electric components, essentially doubling the contribution). We are considering high energy cosmic rays, which are predominantly protons, so we consider the two radiation processes for these \footnote{Very high energy electrons also emit Cherenkov radiation in the pulsar field. However, due to the greater rate of synchrotron emission in this case, the window in which Cherenkov radiation is the dominant effect is far narrower.}. This amounts to changing $m_e \rightarrow m_p$ in (\ref{CrossedSynch}). However factors of $m_e$ appearing in the Cherenkov spectrum (through the parameters $\lambda_{\pm}$) and the cut-off are not changed: the nonlinear terms in the Lagrangian (\ref{Lagrangian},\ref{EH_Lagrangian}) and the mass scale in the cut-off (\ref{cut}) are   determined by electron-positron fluctuations in the vacuum. The total power radiated per unit frequency is again determined by (\ref{PowerTotal}), with (\ref{power}).

For radiation from protons, we also need to compare (\ref{PowerTotal}) with the radiation of pions, which subsequently decay into photons. The spectrum for such radiation is given by \cite{ginzburg1965pions}
\begin{align}
\frac{d P_\pi}{d\omega}
=&
\frac{g^2}{\sqrt{3}\pi c}\gamma^{-2}\omega
\int_y^{\infty}
dx
K_{1/3}(x)
,
\label{Pions}
\end{align}
where $y=\frac{2}{3}\frac{\hbar\omega}{m_e}\frac{m_p}{m_e}\frac{B_S}{B}\gamma^{-2}\left[ 1+\left({\hbar\omega}/{\gamma m_\pi}\right)^2\right]^{3/2},$ $m_\pi$ is the pion mass, and $g^2\simeq 14\hbar c$ is the pion-proton coupling strength.

The strongest magnetic fields observed are those produced by rapidly rotating pulsars. These objects have characteristic attributes of mass and radius, which with rotational period determine the typical field strengths produced. There are two broad classes of pulsar, those with a relatively longer rotational period which have magnetic field strengths $B\sim 10^8$ T, and rapidly rotating ``millisecond pulsars'' which have typical field strengths $B\sim 10^4$ T \cite{camilo1994millisecond}. The cut-off energy found through (\ref{cut}) is $\hbar\omega_{\text{max}} \sim (m_e^3 c^4)/(e \hbar B)$. This corresponds to 22.5 MeV for $B=10^8$ T, or 225 GeV for $B=10^4$ T. Since we are interested in high energy gamma rays, we illustrate the results for millisecond pulsars.

\begin{figure}
\centering
    \includegraphics[width=0.4\textwidth,{trim= 0 0 0cm 0},clip]{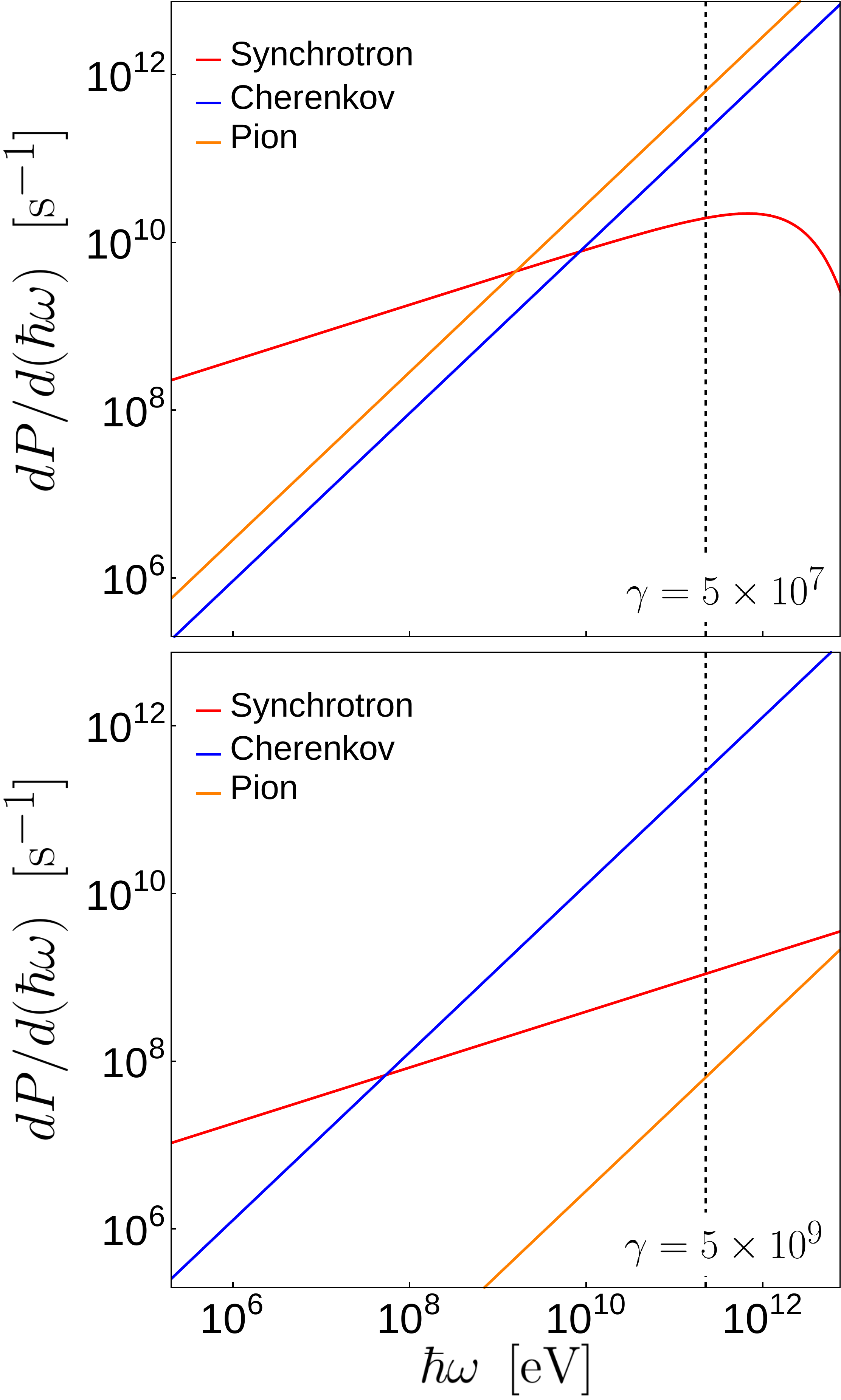}
\caption{
Power radiated via synchrotron, pion and Cherenkov emission, by protons in a magnetic field $B = 10^4$ T, with Lorentz factor $\gamma = 5 \times 10^7$ (upper panel) and $\gamma = 5 \times 10^9$ (lower panel). The cut-off energy is $\hbar \omega_{\text{max}} \simeq 225$ GeV.}
\label{fig:Magnetic_B4}
\end{figure}

Figure~\ref{fig:Magnetic_B4} shows the spectra for Cherenkov, synchrotron and pion radiation for a proton moving perpendicularly to a magnetic field $B = 10^4$ T, for $\gamma=5\times 10^7$ (just above the Cherenkov threshold) and $\gamma=5\times 10^{9}$. For clarity we include only the total Cherenkov contribution.   For $\gamma = 5\times 10^7$, the Cherenkov radiation exceeds synchrotron emission for photon energies above 8.5 GeV, but remains below the pion emission up to $\hbar \omega_\text{max}$. For $\gamma=5\times 10^{9}$, however, Cherenkov radiation is by far the dominant emission channel for photon energies from 54~MeV up to the cut-off. So for the highest energy proton cosmic rays, the highest energy radiation is completely dominated by the Cherenkov process.

There is currently a debate within the astrophysics community concerning the origin of observed excesses of high energy photons found in recent data. For example, observations of intense gamma rays from the Galactic Center \cite{fermi-LAT2017} have prompted a range of possible explanations, such as dark matter annihilation \cite{hooper2011} and unresolved pulsar sources \cite{gordon2015}. The Cherenkov process detailed in this Letter provides a new, and so far unexplored, gamma-ray production mechanism, which we believe warrants further study in this context.

To summarize, in this Letter we have provided a comprehensive, quantitative study of the Cherenkov effect in nonlinear theories of vacuum electrodynamics. This effect---expected due to the reduced phase velocity of light predicted by these theories in regions of strong fields---may provide an alternate radiation mechanism for very high energy particles. We considered two examples of background field with relevance to future experimental or observational campaigns, and determined the possibility of observing Cherenkov radiation in each case. When the background field is a constant crossed field (approximating a laser pulse), the availability of high energy particles appears to put observation of Cherenkov radiation out of reach. In contrast, astrophysics provides environments in which the vacuum Cherenkov effect may be observed, due to the presence of very high energy cosmic rays and strong magnetic fields. We have demonstrated regimes in which radiation due to the nonlinear Cherenkov effect dominates over radiation produced through synchrotron and pion emission, generating very high energy photons. A notable excess of gamma rays with energies in the GeV--TeV range has been observed in various astrophysical contexts, and the vacuum Cherenkov process could provide an alternate explanation for their origin, not previously considered in the literature.

\begin{acknowledgments}
{\em Acknowledgements}---We would like to thank other members of the ALPHA-X Collaboration for useful discussions. This work was supported by the UK EPSRC (Grant No. EP/N028694/1) and a University of Strathclyde DTP studentship. All of the results can be fully reproduced using the methods described in the paper.
\end{acknowledgments}

\bibliographystyle{apsrev4-1}
%


\clearpage

\section{Appendix A: Polarization 3--vectors}
\label{sec:pol}

Here, we derive expressions for the overlap functions $|\bm{\hat{\epsilon}_0}.\bm{\hat{\epsilon}_{\pm}}|^2$ appearing in the Cherenkov spectrum. We begin by orienting our coordinate system with the $\bf{\hat{z}}$-axis along $\bm{\beta}$, so that the Cherenkov angle coincides with the usual polar angle of the emitted radiation, and we have
\begin{equation}
{\bf\hat{k}}=\sin\theta_C\cos\phi{\bf\hat{x}}+\sin\theta_C\sin\phi{\bf\hat{y}}+\cos\theta_C{\bf\hat{z}}.
\end{equation}

ICR is polarized in the $(\bm{\hat{\beta}},{\bf\hat{k}})$ plane perpendicular to ${\bf\hat{k}}$, which together with unit normalization gives
\begin{equation}
\bm{\hat{\epsilon}_0}=\frac{{\bf\hat{z}}-\cos\theta_C\bf{\hat{k}}}{\sin\theta_C}.
\label{epsilon0}
\end{equation}

We now need expressions for $\bm{\hat{\epsilon}_\pm}$. The radiation field tensor is written
\begin{equation}
f_{\mu\nu}=\left( k_\mu a_\nu-k_\nu a_\mu\right)e^{i\varphi},
\label{radiation}
\end{equation}
with the polarization 4-vectors $a^\mu$ given by
\begin{equation}
a^\mu_+= \FudC{\mu}{\nu} k^\nu, \qquad a^\mu_-= \FudCt{\mu}{\nu} k^\nu.
\label{a1}
\end{equation}
To interpret the spatial components of $a^\mu_\pm$ as polarization 3-vectors, their temporal components must vanish, $a^0_\pm=0$, i.e., we must be in the Weyl gauge. This is not in general the case: a background electric field will give rise to a nonzero $a^0_+$, while a background magnetic field generates a nonzero $a^0_-$:
\begin{equation}
a^0_+={\bf k}.\bm{\mathcal E}, \quad a^0_-=-{\bf k}.\bm{\mathcal B},
\end{equation}
where $\bm{\mathcal E}$ ($\bm{\mathcal B}$) is the background electric (magnetic) field.

However, the field (\ref{radiation}) is invariant under the gauge transformation $a^\mu_\pm\rightarrow a^{\prime\mu}_\pm =a^\mu_\pm +{\cal C}_\pm k^\mu$, and choosing ${\cal C}_+=-{\bf k}.\bm{\mathcal E}/\omega$, and similarly for ${\cal C}_-$, the new polarization 4-vectors $a^{\prime\mu}_\pm$ are in the Weyl gauge. We then take $\bm{\hat{\epsilon}_\pm}$ to be the unit normalized 3-vector proportional to the spatial parts of $a^{\prime\mu}_\pm$,
\begin{equation}
\bm{\hat{\epsilon}_+}=\frac{\bm{\mathcal E}+v^{-1}_p{\bf\hat{k}}\times \bm{\mathcal B}- v^{-2}_p(\bm{{\mathcal E}}.{\bf\hat{k}}){\bf\hat{k}}}{{\bm{|}}\bm{\mathcal E}+v^{-1}_p{\bf\hat{k}}\times \bm{\mathcal B}- v^{-2}_p(\bm{{\mathcal E}}.{\bf\hat{k}}){\bf\hat{k}}\bm{|}}.
\label{epsilon+}
\end{equation}
The phase velocity in (\ref{epsilon+}) can be obtained either from the definition of the Cherenkov angle, $v_p=\beta\cos\theta_C$, or from the dispersion relation, $v^2_p=1+2\lambda_+ \FuC{\mu}{\lambda}\FudC{\nu}{\lambda}\hat{k}_\mu\hat{k}_\nu$.

$\bm{\hat{\epsilon}_-}$ is (\ref{epsilon+}) with the substitution $(\bm{\mathcal E},\bm{\mathcal B})\rightarrow (-\bm{\mathcal B},\bm{\mathcal E})$.


\subsection{Overlap functions: Constant magnetic field}


In the constant magnetic field oriented in the $\bf{\hat{y}}$-direction, $a^0_+=0$, so we have
\begin{equation}
\bm{\hat{\epsilon}_+}= |\bm{a}_+|^{-1}\bm{a}_+= \frac{\cos\theta_C{\bf\hat{x}} -\sin\theta_C\cos\phi{\bf\hat{z}}}{\sqrt{1-\sin^2\theta_C \sin^2\phi}}.
\label{B+epsilon}
\end{equation}

The polarization vectors (\ref{epsilon0}) and (\ref{B+epsilon}) can now be combined to give the overlap
\begin{equation}
|\bm{\hat{\epsilon}_0}.\bm{\hat{\epsilon}_+}|^2=\frac{\cos^2\phi}{1-\sin^2\theta_C\sin^2\phi}.
\end{equation}

For the second polarization, $a^0_-=-B|{\bf k}|\sin\theta_C\sin\phi$, so we must take ${\cal C}_-=v_p^{-1}B\sin\theta_C\sin\phi$. Hence we have
\begin{align}
\nonumber \bm{\hat{\epsilon}_-}&=|\bm{a}_- + {\cal C}_-{\bf k}|^{-1}\left(\bm{a}_- + {\cal C}_-{\bf k}\right)\\
&= \frac{\sin\theta_C\sin\phi{\bf\hat{k}}-v^2_p{\bf\hat{y}}}{\sqrt{v^4_p-(2v^2_p-1)\sin^2\theta_C\sin^2\phi}}
\label{B-epsilon}
\end{align}

The polarization vectors (\ref{epsilon0}) and (\ref{B-epsilon}) can now be combined to give the overlap
\begin{align}
\nonumber |\bm{\hat{\epsilon}_0}.\bm{\hat{\epsilon}_-}|^2&=\frac{v^4_p\cos^2\theta_C\sin^2\phi}{v^4_p-(2v^2_p-1)\sin^2\theta_C\sin^2\phi}\\
&= \frac{\cos^2\theta_C\sin^2\phi}{1-\sin^2\theta_C\sin^2\phi}+{\cal O}(\lambda^2_-),
\end{align}
where in the last line we have used $v^2_p=1+{\cal O}(\lambda_-)$.

\section{Appendix B: Validity of the rectilinear motion approximation}
\label{sec:rectilinear}

Here, we demonstrate that the approximation of rectilinear motion is valid for the important features of the Cherenkov spectrum in the examples considered in the Letter. Since no motion is perfectly rectilinear, we assume the particle can turn up to some angle $\theta_\text{max}\ll 1$ and still be considered to move in a straight line.

A proton with Lorentz factor $\gamma\gg 1$ in a magnetic field of strength $B$ undergoes cyclotron oscillations with radius $R=\gamma m_pc/eB$. Approximating its speed as $c$, and that of the emitted radiation as $v_p=c(1-\lambda_\pm B^2)$, in the emission of one wavelength $\lambda$ the proton travels a distance $d=\lambda/\lambda_\pm B^2$. During this emission, then, the proton deviates from rectilinear motion by an angle $\theta=d/R=\lambda e/\gamma m_pc\lambda_\pm B$.

The requirement $\theta<\theta_\text{max}$ implies that the proton may be considered to move in a straight line while emitting radiation of wavelength
\begin{equation}
\label{lambdamax}
\lambda<\lambda_\text{max}=\lambda_+ B\frac{m_pc}{e}\gamma\theta_\text{max} \simeq 1.7\times 10^{-19}\gamma\theta_\text{max}\text{ m},
\end{equation}
where we have chosen $\lambda_+$ as it is more restrictive than $\lambda_-$, and we have used $B=10^4\text{ T}$ as in the Letter.

In terms of the energy of the emitted photon, (\ref{lambdamax}) corresponds to
\begin{equation}
\hbar \omega>\hbar\omega_\text{min}=\frac{7.4\times 10^{12}}{\gamma\theta_\text{max}}\ \text{eV}.
\end{equation}
With $\gamma=5\times 10^7$ (the smallest value considered in the Letter) and assuming a tolerance of $\theta_\text{max}=10^{-2}\text{ rad}$, this corresponds to $\hbar \omega_\text{min}=15\text{ MeV}$. Due to the frequency dependence of the spectrum, the vast majority of radiation satisfies $\omega>\omega_\text{min}$: the ratio of the power radiated in the range $\omega_\text{min}<\omega<\omega_\text{max}$ to the total power radiated assuming rectilinear motion is 
\begin{equation}
 {\cal R}= 1- \frac{\omega^2_\text{min}}{\omega^2_\text{max}}\simeq 1-(4\times 10^{-9}).
\end{equation}
For the other example considered, the range of frequencies over which the rectilinear approximation is valid increases considerably.

\end{document}